\documentclass{article}
\usepackage{amssymb}
\usepackage{amsfonts}

\input{tcilatex}
\begin{document}

\title{Kolmogorovian versus non-Kolmogorovian probabilities in contextual
theories }
\author{Claudio Garola \\
%EndAName
Department of Mathematics and Physics, University of Salento\\
Via Arnesano, 73100 Lecce, Italy\\
e-mail: garola@le.infn.it}
\maketitle

\begin{abstract}
Most scholars maintain that quantum mechanics (QM) is a contextual theory
and that quantum probability does not allow an epistemic (ignorance)
interpretation. By inquiring possible connections between contextuality and
non-classical probabilities we show that a class $\mathbb{T}^{\mu \mathcal{MP%
}}$ of theories can be selected in which probabilities are introduced as
classical averages of Kolmogorovian probabilities over sets of (microscopic)
contexts, which endows them with an epistemic interpretation. The conditions
characterizing $\mathbb{T}^{\mu \mathcal{MP}}$ are compatible with classical
mechanics (CM), statistical mechanics (SM) and QM, hence we assume that
these theories belong to $\mathbb{T}^{\mu \mathcal{MP}}$. In the case of CM
and SM this assumption is irrelevant, as all notions introduced in them as
members of $\mathbb{T}^{\mu \mathcal{MP}}$\ reduce to standard notions. In
the case of QM it leads to interpret quantum probability as a derived
notions in a Kolmogorovian framework, explains why it is non-Kolmogorovian
and provides it with an epistemic interpretation. These results were
anticipated in a previous paper but are obtained here in a general framework
without refererring to individual objects, which shows that they hold even
if only a minimal (statistical) interpretation of QM is adopted to avoid the
problems following from the standard quantum theory of measurement.\medskip 

\textbf{Key words}: Contextuality, non-Kolmogorovian probabilities, quantum
probability, quantum measurements.
\end{abstract}

\section{Introduction}

Probability enters quantum mechanics (QM) via Born's rule and is usually
interpreted in terms of frequencies of the outcomes obtained when
measurements are performed. However, it turns out to be non-Kolmogorovian,
in the sense that the probability measures associated with quantum states do
not satisfy the assumptions of Kolmogorov's probability theory. In
particular, the set of events for every probability measure associated with
a quantum state is the orthomodular lattice of \textit{standard quantum logic%
} (QL), which is nondistributive (except for some special cases), at
variance with the set of events in Kolmogorov's theory, which is a Boolean
lattice.

The standard interpretation of QM introduces another nonclassical feature of
QM, i.e. the doctrine that, whenever a physical system in a given state is
considered, a quantum observable generally has not a prefixed value but only
a set of \textit{potential} values, and that a measurement on an individual
example of a physical system (\textit{individual object} in the following)
actualizes one of these values, yielding an outcome that depends on the
specific measurement procedure that is adopted (\textit{contextuality}).
This doctrine is usually maintained to be proven correct by some "no-go"
theorems that supply a mathematical support to the standard (Copenhagen)
interpretation of QM and show, in particular, that contextuality occurs also
in the case of measurements on far-away subsystems of a composite physical
system (\textit{nonlocality}). However, maintaining that QM deals with
individual objects and their properties raises the "measurement problem" of
QM (see, e.g., Busch et al., 1996), which is still considered unsolved by
many scolars and implies known paradoxes.\footnote{%
There is a huge literature on these topics, which goes back to the early
days of QM. We limit ourselves here to recall that the EPR and the QL issues
were started by the famous papers by Einstein, Podolski and Rosen (1935) and
by Birkhoff and von Neumann (1936), respectively, while the nonlocality and,
more generally, the contextuality of QM were accepted by most physicists as
"mathematically proven" after the publication of Bell's (1964, 1966) and
Kochen-Specker's (1967) theorems (later supported by numerous different
proofs of the same or similar theorems, among which, in particular, the
proof of nonlocality provided by Greenberger, Horne, Shimony and Zeilinger,
1990, which does not resort to inequalities).}

Because of contextuality, it is a widespread belief that quantum probability
does not admit an epistemic interpretation (the term \textit{epistemic} is
meant here in a broad sense, i.e., as referring to our degree of
knowledge/lack of knowledge). Indeed, generally one cannot consider a
property of a quantum system as either possessed or not possessed by the
system independently of any measurement, even if the state of the system is
known. Hence one cannot look at the values of the probability measure
associated with the state as indexes of the degree of ignorance of the
properties possessed by the system. Probability should rather be seen as an
intrinsic feature of the microworld, i.e., it is \textit{ontic}.

The standard view expounded above is obviously legitimate, but further
investigation on possible links between quantum probability and
contextuality may suggest alternative views. We have inquired into such
links in a previous paper (Garola, 2018), where both macroscopic contexts
and microscopic contexts ($\mu $\textit{-contexts}) were introduced and
quantum probability measures were interpreted as classical averages of
Kolmogorovian probability measures on $\mu $-contexts.\footnote{%
A valuable "contextual approach to quantum formalism" has been provided by
Khrennikov (2009a, 2009b) in the framework of the "V\"{a}xj\"{o} school".
This approach, however, is basically different from ours. Khrennikov
considers indeed contexts "as a generalization of a widely used notion of 
\textit{preparation procedure}" (2009b), which includes also \textit{%
selection procedures} that are \textit{registration procedures} in the sense
of Ludwig (1983). In our approach, instead, macroscopic measurement
procedures are associated with macroscopic measurement contexts, which seems
to be compatible with Khrennikov's view, but an essential role is played by
the $\mu $-contexts underlying macroscopic measurement contexts, which are
not considered by Khrennikov.} At the best of our knowledge, our approach is
innovative, as it focuses on an analysis and a rational reconstruction of
the basic language of QM, thus adopting a methodology that is typical of
analytic philosophy but rather unusual in physics. Yet the language worked
out in the paper quoted above is a first order predicate calculus in which
an individual variable $x$ is interpreted on individual objects. Hence our
results rest on an interpretation of QM that is not free of problems and
paradoxes. We propose in the present paper a more general view, singling out
a class of theories, including classical mechanics (CM), statistical
mechanics (SM) and QM, in which non-Kolmogorovian probability measures are
introduced as derived notions in a Kolmogorovian framework, taking into
account contextuality but making no reference to individual objects. Let us
therefore summarily describe our procedures.

First of all, after some epistemological and physical preliminaries
(Sections 2 and 3, respectively), we consider in Section 4 a class $\mathbb{T%
}$ of theories in which the basic notions of \textit{physical system} (or 
\textit{entity}), \textit{state}, \textit{property} and \textit{macroscopic
context} are introduced, and then work out a propositional language $L$
that, for every $\mathcal{T}\in \mathbb{T}$ in which every macroscopic
context can be associated with a set of $\mu $\textit{-}contexts, formalizes
a fragment of the natural language expressing basic features of the entities
considered in $\mathcal{T}$. The set of elementary (or \textit{atomic})
propositions of $L$ is partitioned into a subset of \textit{atomic state
propositions} and a subset of \textit{atomic context-depending propositions}%
. A proposition of the former subset affirms that an entity $H$ of $\mathcal{%
T}$ is in a given state. A proposition of the latter subset affirms that an
entity $H$ of $\mathcal{T}$ possesses a given property in a given $\mu $%
-context (we stress that no\ atomic proposition assigning a property of $H$\
without referring to a $\mu $-context exists in $L$)$.$ Then we select a
subclass $\mathbb{T}^{\mu \mathcal{MP}}\subset \mathbb{T}$\ by introducing a
classical probability measure on the set of all (atomic and molecular)
propositions of $L$ (Section 5) and a family of classical probability
measures defined on subsets of $\mu $-contexts (Section 6), each element of
the family corresponding to a \textit{measurement procedure} that determines
a \textit{macroscopic measurement context} associated with a property. We
can thus define a notion of \textit{compatibility} in the set of all
properties of $L$, hence a notion of \textit{testability} in the set of all
propositions of $L$, and use the foregoing classical probability measures
conjointly to define the notion of \textit{mean conditional probability} on
the subset of all testable propositions, together with the related notion of 
\textit{mean probability test}. Hence mean conditional probabilities admit
an \textit{epistemic} interpretation, but are not bound to satisfy
Kolmogorov's axioms, because they are obtained by averaging over classical
probability measures.

Based on the definitions and results expounded above we focus (Section 7) on
the set $\mathcal{E}$\ of all properties, on which a family of mappings $%
\mathcal{E\longrightarrow }\left[ 0,1\right] $ can be introduced by means of
mean conditional probabilities, parametrized by the set $\mathcal{S}$\ of
all states. This family induces a preorder relation $\prec $ on $\mathcal{E}$%
. We show that, if suitable structural conditions are satisfied, each
element of the family is a \textit{generalized probability measure} (or 
\textit{q-probability}) on $(\mathcal{E,}\prec )$, which reduces to a
classical probability measure whenever $(\mathcal{E,}\prec )$ is a Boolean
lattice. Generalized probability measures can be empirically tested and
admit an epistemic interpretation, but generally do not satisfy Kolmogorov's
assumptions. Moreover, they allow the definition of a new kind of
conditioning that refers to a sequence of measurements and is conceptually
different from classical conditioning.

We are thus ready to discuss the implications of our framework in the
special cases of CM, SM and QM. To this end, we firstly show that the
characterizing features of $\mathbb{T}^{\mu \mathcal{MP}}$ are compatible
with these theories (Section 8), hence we assume that CM, SM and QM belong to%
$\mathbb{T}^{\mu \mathcal{MP}}$.\footnote{%
There are nowadays also nonphysical theories that can be maintained to
belong to $\mathbb{T}^{\mu \mathcal{MP}}$, as the models in cognitive
sciences that use a quantum formalism (see, e.g., Aerts et al., 2015, 2016).
We do not deal with this issue in the present paper for the sake of brvity.}
Then, we show that this assumption has a great explanatory power. Indeed,
leaving apart SM for the sake of brevity, we show in Section 9 that all the
notions introduced above collapse into standard notions in CM, consistently
with the non-contextual character of this theory. Moreover, by considering
QM in Section 10 we recover the following results that have been anticipated
in the paper mentioned above (Garola, 2018).

(i) The probability measures on the set of properties induced by the Born
rule in QM can be considered as the specific form that q-probabilities take
in QM. Hence they are interpreted as derived notions within a Kolmogorovian
framework and their non-classical character can be explained in classical
terms. This explanation implies that quantum probability can be provided
with an epistemic rather than an \textit{ontic }interpretation by taking
into account $\mu $-contexts.

(ii) The relation of compatibility on the set of all physical properties
that occurs in QM can be considered as a specific form of the relation of
compatibility introduced in our general framework.

(iii) The conditional probability usually introduced in quantum probability
can be considered as the specific form of the new kind of conditioning
introduced in our general framework.

These results are now obtained without referring to individual objects (no
individual variable occurs in $L$), hence they hold even if only a minimal
(statistical) interpretation of QM is accepted (see, e.g., Ballentine, 1970;
Busch et al., 1996) that avoids the problems raised by the standard quantum
theory of measurement.

Finally, we close our paper with some conclusive remarks (Section 9), and
then add an Appendix. This addition is motivated by the fact that the
notions of mean conditional probability and mean probability test are
conceptually similar to the notions of \textit{universal average} and 
\textit{universal measurement}, respectively, introduced by Aerts and
Sassoli de Bianchi (see, e.g., 2014, 2017). In particular, in the case of QM
our approach provides a description of the measurements testing
probabilities that recalls the proposal of these authors.\footnote{%
We stress that our general framework is not a hidden variables theory for
QM, at least in a standard sense. Indeed, $\mu $-contexts are associated
(generally many-to-one) with macroscopic measurement procedures, not with
states or properties of the entity that is being measured. Our perspective
reminds instead Aerts' hidden measurements approach (1986).} But there are
also some relevant differences between the two approaches. In particular,
quantum probability is considered as ontic by Aerts and Sassoli de Bianchi,
while we prove in the present paper that it also admits an epistemic
interpretation. Our Appendix therefore aims to provide a brief and intuitive
account of the aforesaid similarities and differences.

\section{Epistemological preliminaries}

According to the epistemological view called \textit{standard
epistemological conception}, or \textit{received view} (see, e.g.,
Braithwaite, 1953; Nagel, 1961; Hempel, 1965; Carnap, 1966), a
fully-developed physical theory $\mathcal{T}$ is in principle expressible by
means of a metalanguage in which a \textit{theoretical language} $L_{T}$, an 
\textit{observational language} $L_{O}$ and \textit{correspondence} (or 
\textit{epistemic})\textit{\ rules} $R_{C}$ connecting $L_{T}$ and $L_{O}$
can be distinguished. The theoretical apparatus of $\mathcal{T}$, expressed
by means of $L_{T}$, includes a \textit{mathematical structure} and,
usually, an \textit{intended interpretation} which is a \textit{direct} and 
\textit{complete} physical model of the mathematical structure (this model
is often anticipated by the choice of the nouns of the theoretical terms and
it is not indispensable in principle, but plays a fundamental role in the
intuitive comprehension, justification and development of the theory;\
think, e.g., to the trajectories of point-like particles in CM or to the
geometrical representation of electromagnetic waves). The observational
language $L_{O}$\ describes instead an empirical domain, hence it has a
semantic interpretation, so that the correspondence rules $R_{C}$\ provide
an \textit{empirical interpretation} of the mathematical structure. Such an
interpretation, however, is often complicated and/or problematic (e.g.,
one-dimensional orthogonal projection operators may represent both a pure
state and a dichotomic observable in QM, i.e., different physical entities).
Moreover, it is generally \textit{indirect}, in the sense that there are
theoretical entities that are connected with the empirical domain only via 
\textit{derived} theoretical entities, and \textit{incomplete}, in the sense
that only limited ranges of values of the theoretical entities are
interpreted (e.g., self-adjoint operators correspond in QM to measuring
apparatuses whose outcomes match the eigenvalues of the operators only in
finite intervals of the real axis).

The received view has been criticized by some authors (see, e.g. Kuhn, 1962;
Feyerabend, 1975) and is nowadays maintained to be outdated by several
scholars. Nevertheless, we deem that its basic ideas are still
epistemologically relevant. In particular, this view led us to focus our
attention on the languages of physical theories, suggesting to explore their
similarities and differences by analysing their syntax and semantics to find
out the roots of several open problems in the foundations of such theories.
The results that we have obtained following that suggestion are sometimes
unexpected and challenge well established beliefs (see, e.g., Garola and
Sozzo, 2013; Garola et al., 2016; Garola, 2017; Garola, 2018).

We add that in the standard language of physical theories the distinctions
introduced by the received view are usually overlooked, and the various
linguistic components are mixed together (e.g., the term \textquotedblleft
observable\textquotedblright\ may denote in QM a self-adjoint operator on a
Hilbert space $\mathcal{H}$, and in this sense it belongs to $L_{T}$, but
also a physical entity associated with a set of measurement procedures, and
in this sense it belongs to $L_{O}$; the term \textquotedblleft
state\textquotedblright\ may denote a vector of $\mathcal{H}$, but also a
physical entity associated with a set of preparing procedures; etc.). Only a
rational reconstruction of the language of a theory can lead to clearly
distinguish the various elements that occur in it according to the received
view. For the sake of simplicity we therefore retain here some of the basic
ideas of such view that we consider epistemologically relevant, but adopt a
simpler scheme. To be precise, we maintain that every advanced scientific
theory $\mathcal{T}$ is expressible by means of a fragment of the natural
language enriched with technical terms and is characterized by a pair $(F,I)$%
, with $F$ a logical and mathematical formalism that may have an intended
interpretation and $I$ an \textit{empirical interpretation}, indirect and
incomplete in the sense explained above, that establishes connections
between $F$ and an empirical domain. Moreover, in some locutions (as
\textquotedblleft the minimal interpretation of QM\textquotedblright , etc.)
the interpretation $I$ will be distinguished from the theory, following a
standard use.

\section{Physical preliminaries}

The main ideas for our general treatment are suggested by a typical case of
contextual theory, i.e. QM. Therefore we consider this theory in the present
section. For the sake of intuitivity we refer here to an interpretation of
QM that is "realistic" in the sense thart it assumes that QM deals with
individual objects and their properties (see, e.g., Busch et al., 1996),
even if our general theory avoids referring to individual objects, as
anticipated in Section 1. Moreover, we adopt a standard physical language in
which the distinctions emphasized\ in Section 2 are not explicitly
introduced.

First of all we recall that in most presentations of QM the notions of 
\textit{physical system}, or \textit{entity}, \textit{(physical) property}
and \textit{(physical) state} are fundamental, and that, according to some
known approaches to the foundations of QM (see, e.g., Beltrametti and
Cassinelli, 1981; Ludwig, 1983), states are empirically interpreted as
classes of probabilistically equivalent \textit{preparation procedures}, or 
\textit{preparing devices}, and properties as classes of probabilistically
equivalent \textit{dichotomic (yes-no) registering devices}. This
interpretation suggests an intuitive explanation of the fact that QM yields
only probabilistic predictions. Indeed, one can adopt a picture according to
which a microscopic world underlies the macroscopic world of our everyday
experience and note that there are two possible sources of randomness for
the outcomes of a measurement, as follows (see also Khrennikov, 2015).

(i) When an individual object is prepared by activating a preparation
procedure associated with a state $S$, we control only macroscopic
variables, not the physical situation at a microscopic level. Thus different
individual objects produced by the preparation procedure are not bound to
yield the same outcomes.

(ii) When a registering device is activated to perform a measurement, many 
\textit{microscopic contexts} can be associated with it, and different
microscopic contexts that we cannot control may affect in different ways the
outcome of the measurement.

The picture above, however, does not distinguish QM from SM. This crucial
distinction can be established as follows. Consider QM and a preparation
procedure $\pi $ in the class $S$. When activated, $\pi $ produces an
individual object $x$ (which can be identified with the act of activation
itself if one wants to avoid ontological commitments). Hence, after the
activation a sentence that affirms that $x$ is in the state $S$ is \textit{%
true} and a sentence that affirms that $x$ is in a state $S^{\prime }\neq S$
is \textit{false}. Then, given an individual object $x$ in the state $S$,
activating a registering device $r$ in the class $E$ performs a test of the
property $E$, but the result of the test generally depends on the set of
properties (pairwise compatible and compatible with $E$) that are tested
together with $E$. It follows in fact from some known proofs of Bell's and
Kochen-Specker's theorems mentioned in Section 1 (see, e.g., Greenberger et
al., 1990; Mermin 1993) that, if the laws of QM have to be preserved in
every conceivable physical situation, the outcome that is obtained depends
on the set of the registering devices that are activated together with $r$,
i.e., on the macroscopic context $C_{m}$\ determined by the whole
(macroscopic) \textit{measurement} $m$ that is performed. Briefly, QM is a 
\textit{contextual} theory, at variance with SM.

Contextuality means that it is impossible in QM to assign a truth value to a
sentence stating that $x$ has (or \textit{possesses}) a property $E$
disregarding the measurement context. In other words, the natural everyday
language and the technical language of classical physics, whose elementary
sentences state properties of individual\ objects independently of any
observation, are unsuitable for QM (which is the source of most "quantum
paradoxes" in our opinion). This fundamental feature of QM was clearly
implicit in Bohr's holistic view (see, e.g., Bohr, 1958) or in Heisenberg's
distinction between "potential" and "actual" properties (see, e.g.,
Heisenberg, 1958), but it was maintained to be definitively "mathematically
proven" only after the statement of Bell's and Kochen-Specker's theorems
quoted above.\footnote{%
We have emphasized in some previous papers (see, e.g., Garola, 1999; Garola
and Pykacz, 2004; Garola and Sozzo, 2010; Garola and Persano, 2014) that the
epistemological clause "the laws of QM have to be preserved in every
conceivable physical situation" is essential in the proofs of Bell's and
Kochen-Specker's theorems. Nevertheless, this clause generally is not
explicitly noticed or stated, possibly because it seems to be unquestionably
justified by the outstanding success of QM. Yet it must be observed that all
the proofs mentioned above proceed \textit{ab absurdo}, considering physical
situations in which noncompatible physical properties are assumed to be
simultaneously possessed by an individual object and showing that this
assumption leads to contradictions with well established quantum laws. But
in the aforesaid situations the quantum laws that are applied can never be
simultaneously tested, hence hypothesizing that they hold anyway seems more
consistent with a classical than with a quantum view. One can therefore try
to give up the aforesaid clause, but then the proofs of Bell's and
Kochen-Specker's theorems cannot be completed. This conclusion opens the way
to the attempt at recovering noncontextual interpretations of QM (see, e.g.,
Garola, 2015; Garola et al., 2016). We, however, adhere to the standard view
in the present paper, even if our arguments apply to every theory in which
contexts can be defined, irrespective of whether the results of measurements
are context-depending (locally, or also at a distance) or not.}

At first sight one can think that a possible answer to the problems raised
by the contextuality of QM is assuming that the basic language of QM is the
nonstandard logic of quantum propositions introduced by Birkhoff and von
Neumann (1936), which implies a nonclassical notion of truth (\textit{%
quantum truth}) according to some authors (see, e.g., R\'{e}dei, 1998; Dalla
Chiara et al., 2004). But this answer does not grasp the point in our
opinion. Indeed, we have proven in some previous papers that quantum logic
can be embedded (preserving the order but not the algebraic structure) into
a classical logic (Garola, 2008; Garola and Sozzo, 2013) or into a pragmatic
extension of classical logic (Garola, 2017). Moreover, these results are
supported by some former results that show that there are examples of
classical macroscopic systems that exhibit a quantum structure (see, e.g.,
Aerts, 1999).

On the other side, one can maintain that contextuality, implying a breakdown
with a classical view of the world, is a fundamental feature that should be
incorporated into the basic language of QM rather than recognized at a later
stage. By associating this idea with the above picture of the sources of
randomness in QM, we observe that, generally, the macroscopic context $C_{m}$
determined by a macroscopic measurement $m$ may be produced by many
different microscopic physical situations that cannot be distinguished at a
macroscopic level (though they can be described, in principle, by QM
itself). Hence we can associate $C_{m}$ with a set $\mathcal{C}_{m}$ of 
\textit{microscopic contexts} ($\mu $\textit{-contexts}; of course, $%
\mathcal{C}_{m}$ could reduce to a singleton in special cases) and then
assume that the truth value of a sentence asserting that $x$ possesses the
property $E$ generally depends on $m$ through the $\mu $-context that is
realized when $m$ is performed. But we cannot know this $\mu $-context,
hence only a probability of it can be given which expresses our degree of
ignorance of it (we naively argue here as though the set $\mathcal{C}_{m}$
were discrete, to avoid technical complications).

Summing up, our picture leads us to conclude that a truth value can be
supposed to exist which is consistent with QM only in the case of a sentence
asserting that an individual object $x$ possesses a property $E$ in a given $%
\mu $-context $c$, not in the case of a sentence simply asserting that $x$
possesses a property $E$. Moreover, in general this value cannot be deduced
from the laws of QM, which are probabilistic laws that make no explicit
reference to contexts: hence, we generally do not know it.

The conclusions above have an important consequence. Every quantum
prediction concerns probabilities, hence in our present perspective testing
it requires evaluating frequencies of outcomes. A typical test of this kind
consists in preparing a broad set of individual objects in a given state $S$
and then performing on each object the same macroscopic measurement $m$ by
activating one or more (compatible) registering devices. The macroscopic
context $C_{m}$ then is the same for every individual object, but the $\mu $%
-context $c\in \mathcal{C}_{m}$ generally changes in an unpredictable way.
Thus we meet two distinct sources of randomness. The first is the state $S$,
be it a pure state or a mixture, (see (i) above; note that we could
introduce $\mu $-contexts also referring to the preparation procedures
associated with $S$ by the empirical interpretation, but this would
uselessly complicate our framework). The second is the unpredictable change
of the $\mu $-context that occurs when performing $m$ on different
individual objects (see (ii) above). Because of the former source we would
generally obtain different results when iterating $m$, even if we could fix
the $\mu $-context $c\in \mathcal{C}_{m}$, so that for every property $E$ we
could evaluate a frequency approaching (in the large numbers limit) the
probability that an individual object in the state $S$ possesses the
property $E$ in the $\mu $-context $c$. Because of the latter source we can
only assign a probability to every $c\in \mathcal{C}_{m}$ and conclude that
the frequencies that are obtained actually approach the mean over $\mathcal{C%
}_{m}$\ of the foregoing probabilities. It is then reasonable to identify
this mean with the quantum probability of $E$ in the state $S$, which
implies that quantum probabilities take simultaneously into account both the
sources of randomness listed above. We will see in the next sections that
this idea, together with contextuality, can explain the non-Kolmogorovian
character of quantum probabilities, together with the rather surprising fact
that their values neither depend on $\mu $-contexts nor on macroscopic
contexts (see, e.g., Mermin, 1993). To avoid unnecessary restrictions of our
framework, however, we will not refer in the following to individual objects
and consider only measurements directly testing probabilities, consistently
with the minimal interpretations of QM (see Section 1).

\section{The classical propositional language $L$}

Bearing in mind the epistemological and physical preliminaries in Sections 2
and 3, we introduce the following definition.\medskip

\textbf{Definition} \textbf{4.1.} \textit{We denote by }$\mathbb{T}$\textit{%
\ the class of theories in which the notions of }entity\textit{, }property%
\textit{\ and }state\textit{\ are explicitly introduced, together with a
notion of }measurement\textit{\ (hence, implicitly, of }macroscopic context%
\textit{). We then denote by }$\mathbb{T}^{\mu }\subset \mathbb{T}$ \textit{%
the subclass of theories in which each macroscopic context can be associated
with a set of }microscopic contexts\textit{\ (}$\mu $-contexts\textit{%
).\medskip }

We implement now the idea of incorporating contextuality in the basic
language of a theory $\mathcal{T}\in \mathbb{T}^{\mu }$ by constructing a
formalized language $L$ that is intended to provide a rational
reconstruction of the basic language of every $\mathcal{T}\in \mathbb{T}%
^{\mu }$ (hence $L$ can be considered as a part of the formalism of $%
\mathcal{T}$). To this end we agree to use standard symbols in set theory
and logic. In particular, $^{c}$, $\cap $, $\cup $, $\subset $, $\backslash $%
, $\emptyset $ and $%
%TCIMACRO{\U{2119} }%
%BeginExpansion
\mathbb{P}
%EndExpansion
(\Psi )$ will denote complementation, intersection, union, inclusion,
difference, empty set and power set of the set $\Psi $, respectively.
Furthermore $N$ will denote the set of natural numbers.\medskip

\textbf{Definition 4.2.} \textit{We call }entity\textit{\ the triple }$H%
\mathbb{=}(\mathcal{E}$\textit{, }$\mathcal{S}$,\textit{\ }$\mathcal{C})$%
\textit{, where }$\mathcal{E}$\textit{, }$\mathcal{S}$ \textit{and }$%
\mathcal{C}$\textit{\ are disjoint sets whose elements we call }properties%
\textit{, }states\textit{\ and }$\mu $\textit{-}contexts\textit{,
respectively. Then, a basic language }$L$\textit{\ for }$H$\textit{\ is a
classical propositional language, constructed as follows.\smallskip }

Syntax.

\textit{(i) A set }$\Pi _{\mathcal{EC}}^{a}=\left\{ \alpha _{Ec}\mid E\in 
\mathcal{E},c\in \mathcal{C}\right\} $\textit{\ of atomic }%
context-depending\ propositions,\textit{\ a set }$\Pi _{\mathcal{S}%
}^{a}=\left\{ \alpha _{S}\mid S\in \mathcal{S}\right\} $ \textit{of atomic }%
state propositions\textit{\ and a set }$\Pi ^{a}$ $=\Pi _{\mathcal{EC}%
}^{a}\cup \Pi _{\mathcal{S}}^{a}$ \textit{of} atomic\textit{\ propositions.}

\textit{(ii) Connectives }$\lnot $ \textit{(not), }$\wedge $\textit{\ (and), 
}$\vee $\textit{\ (or).}

\textit{(iii) Parentheses }$($\textit{,}$)$\textit{.}

\textit{(iv) A set }$\Pi $\textit{\ of atomic and molecular\ propositions of 
}$L$,\textit{\ obtained by applying recursively standard formation rules in
classical logic (to be precise, for every} $A\in \Pi ^{a}$, $A\in \Pi $%
\textit{; for every} $A\in \Pi $, $\lnot A\in \Pi $\textit{; for every }$%
A,B\in \Pi $\textit{,} $A\wedge B\in \Pi $\textit{\ and }$A\vee B\in \Pi $%
\textit{)}.\smallskip

Semantics.

\textit{A set }$W$\textit{\ of }truth assignments\textit{\ on }$\Pi $\textit{%
, each element of which is a mapping}

\begin{center}
$w:\Pi \longrightarrow $\textit{\ }$\{t,f\}$
\end{center}

\noindent \textit{(where }$t$\textit{\ stands for }true\textit{\ and }$f$%
\textit{\ for }false\textit{) that satisfies the standard (recursive)
assignment rules of classical logic (to be precise, let }$A,B\in \Pi $%
\textit{; then, }$w(\lnot A)=t$\textit{\ iff }$w(A)=f$\textit{, }$w(A\wedge
B)=t$ \textit{iff }$w(A)=t$\textit{\ and }$w(B)=t$\textit{, }$w(A\vee B)=t$ 
\textit{iff }$w(A)=t$\textit{\ or }$w(B)=t$\textit{) and, furthermore, is
such that, for every }$S$\textit{,} $S^{\prime }\in \mathcal{S}$, $S\neq
S^{\prime }$\textit{\ implies that }$w(\alpha _{S^{\prime }})=f$ whenever $%
w(\alpha _{S})=t$.\textit{\smallskip }

We note explicitly that the last clause in the definition of $w$\ is
suggested by the interpretation of states as equivalence classes of
preparation procedures (see Section 3).

The logical preorder and the Lindenbaum-Tarski algebra of $L$ can then be
introduced in a standard way, as follows.\medskip

\textbf{Definition 4.3.} \textit{We denote by }$<$\textit{\ and }$\equiv $%
\textit{\ the (reflexive and transitive) relation of logical preorder and
the relation of logical equivalence on }$\Pi $\textit{, respectively,
defined by standard rules in classical logic (to be precise, for every }$%
A,B\in \Pi $\textit{, }$A<B$\textit{\ iff, for every }$w\in W$\textit{, }$%
w(B)=t$\textit{\ whenever }$w(A)=t$\textit{, and }$A\equiv B$\textit{\ iff }$%
A<B$\textit{\ and }$B<A$\textit{). Moreover we put }$\Pi ^{\prime }=\Pi
/\equiv $\textit{\ and denote by }$<^{\prime }$ \textit{the partial order
canonically induced by }$<$\textit{\ on }$\Pi ^{\prime }$\textit{. Then }$%
(\Pi ^{\prime },<^{\prime })$\textit{\ is a boolean lattice (the
Lindenbaum-Tarski algebra of }$L$\textit{) whose operations }$\lnot ^{\prime
}$\textit{, }$\wedge ^{\prime }$\textit{,}$\vee ^{\prime }$\textit{\ are
canonically induced on }$\Pi ^{\prime }$\textit{\ by }$\lnot $\textit{, }$%
\wedge $\textit{, }$\vee $\textit{, respectively.\medskip }

As stated in Definition 4.2, the language $L$ is a classical propositional
language. Its interpretation, however, introduces some innovative features.
Indeed the words \textit{state}, \textit{property} and $\mu $-\textit{context%
} occur in Definition 4.2 just as nouns of elements of sets, but obviously
refer to an empirical interpretation that makes such elements correspond to
empirical entities denoted by the same nouns. Then, a state $S$ is
associated in $L$ with a state-proposition $\alpha _{S}$ that is argument of
truth assignments and is interpreted as "the entity $H$ is in the state $S$"
(at variance with known views in quantum logic that consider states as 
\textit{possible worlds} of a Kripkean semantics; see, e.g., Dalla Chiara et
al., 2004). A property $E$ is associated instead with a family $\left\{
\alpha _{Ec}\right\} _{c\in \mathcal{C}}$ of context-depending propositions
of $L$, where $\alpha _{Ec}$ is argument of truth assignments and is
interpreted as "the entity $H$ possesses the property $E$ in the $\mu $%
-context $c$".

\section{A $\protect\mu $-contextual probability structure on $L$}

Following Williamson (2002), we introduce now a probability measure on $L$
by means of the following definitions and propositions.\medskip

\textbf{Definition 5.1.}\textit{\ Let }$A\in \Pi $\textit{. Then, we set}

\begin{center}
$Ext:\Pi \longrightarrow 
%TCIMACRO{\U{2119} }%
%BeginExpansion
\mathbb{P}
%EndExpansion
(W),A\longrightarrow \{w\in W|w(A)=t\}$
\end{center}

\noindent \textit{and say that }$Ext(A)$ \textit{is the }extension\textit{\
of the proposition }$A$\textit{.\medskip }

We stress that the extension of a proposition $A$ generally depends on the $%
\mu $-contexts that occur in the formal expression of $A$.\medskip

\textbf{Proposition 5.1.}\textit{\ The mapping }$Ext$\textit{\ satisfies the
following conditions.}

\textit{(i) For every }$A\in \Pi $\textit{, }$Ext(\lnot A)=W\setminus
Ext(A)=(Ext(A))^{c}$\textit{.}

\textit{(ii) For every }$A,B\in \Pi $\textit{, }$Ext(A\wedge B)=Ext(A)\cap
Ext(B)$\textit{.}

\textit{(iii) For every }$A,B\in \Pi $\textit{, }$Ext(A\vee B)=Ext(A)\cup
Ext(B)$\textit{.}

\textit{(iv) For every }$A\in \Pi $\textit{, }$Ext(A\vee \lnot A)=W$\textit{%
\ and }$Ext(A\wedge \lnot A)=\emptyset $\textit{.}

(v) \textit{For every }$A,B\in \Pi $\textit{, }$A<B$\textit{\ iff }$%
Ext(A)\subset Ext(B)$\textit{\ and }$A\equiv B$\textit{\ iff }$Ext(A)=Ext(B)$%
.

\textit{Moreover, the algebraic structure }$\Theta =(Ext(\Pi ),^{c},\cap
,\cup )$\textit{\ is a Boolean algebra isomorphic to }$(\Pi ^{\prime },\lnot
^{\prime },\wedge ^{\prime },\vee ^{\prime })$\textit{.}\smallskip

Proof. Straightforward from Definitions 4.1, 4.2 and 5.1.\medskip

\textbf{Definition 5.2}\textit{. Let }$\Phi =(W,\Theta ,\xi )$\textit{\ be a
classical probability space,\footnote{%
Following a standard terminology, we call \textit{classical probability
space }here any triple $(\Omega ,\Sigma ,\xi )$, where $\Omega $ is a set, $%
\Sigma $ is a Boolean $\sigma $-subalgebra of $%
%TCIMACRO{\U{2119} }%
%BeginExpansion
\mathbb{P}
%EndExpansion
(\Omega )$, and $\xi :\Sigma \longrightarrow \lbrack 0,1]$ is a mapping
satisfying the following conditions: (i) $\xi (\Omega )=1$; (ii) if $\left\{
\Delta _{i}\right\} _{i\in N}$ is a family of pairwise disjoint elements of $%
\Sigma $, then $\xi (\cup _{i}\Delta _{i})=\Sigma _{i}\xi (\Delta i)$.} let }%
$\Pi ^{+}\subset \Pi $\textit{\ be the set of propositions such that, for
every }$B\in \Pi ^{+}$\textit{, }$\xi (Ext(B))\neq 0$\textit{, and let }$p$%
\textit{\ be a binary mapping such that}

\begin{center}
$p:\Pi \times \Pi ^{+}\longrightarrow \lbrack 0,1],(A,B)\longrightarrow
p(A\mid B)=\frac{\xi (Ext(A)\cap Ext(B))}{\xi (Ext(B))}$.
\end{center}

\textit{We say that the pair }$(\Phi ,p)$\textit{\ is a }$\mu $-contextual
probability structure\textit{\ on }$L$\textit{\ and that }$p(A\mid B)$%
\textit{\ is the }$\mu $-contextual conditional probability\textit{\ of }$A\ 
$\textit{given }$B$\textit{. Moreover, whenever }$Ext(B)=W$\textit{\ we say
that }$p(A\mid B)$ \textit{is the }$\mu $-contextual absolute probability%
\textit{\ of }$A$\textit{\ and simply write }$p(A)$\textit{\ in place of }$%
p(A\mid B)$\textit{.\medskip }

The terminology introduced in Definition 5.2 (where the word $\mu $\textit{%
-contextual} emphasizes that the values of $p$ depend on $\mu $-contexts
through the propositions of $L$)\textit{\ }is justified by the following
statement.\medskip

\textbf{Proposition 5.2.} \textit{Let }$B\in \Pi ^{+}$\textit{. Then, the
mapping}

\begin{center}
$p_{B}:\Pi \longrightarrow \lbrack 0,1],A\longrightarrow p(A\mid B)$
\end{center}

\textit{satisfies the following conditions.}

\textit{(i) Let }$A\in \Pi $\textit{\ be such that }$Ext(A)=W$ \textit{%
(equivalently,} $A\equiv A\vee \lnot A$\textit{). Then, }$p_{B}(A)=1$.

\textit{(ii) Let} $\left\{ A_{i}\right\} _{i\in N}$\textit{\ be a family of
propositions such that, for every }$k,l\in N$, $Ext(A_{k})\cap
Ext(A_{l})=\emptyset $\textit{\ (equivalently, }$A_{k}<\lnot A_{l}$\textit{%
). Then, }$p_{B}(\vee _{i}A_{i})=\sum_{i}p_{B}(A_{i})$.\smallskip

Proof. Straightforward.\medskip

Proposition 5.2 shows indeed that, for every $B\in \Pi ^{+}$\textit{, }$%
p_{B} $ is a \textit{probability measure} on $(\Pi ,\lnot ,\wedge ,\vee )$.
Moreover, it obviously implies Bayes' theorem, that is the equation $%
p(B)p(A\mid B)=p(A)p(B\mid A)$.\medskip

\textbf{Remark 5.5.1. }Let $R,S\in \mathcal{S}$ and let $\alpha _{S}\in \Pi
^{+}$. Then, we obtain from Definition 5.2

\begin{center}
$p(\alpha _{R}\mid \alpha _{S})=\frac{\xi (Ext(\alpha _{R})\cap Ext(\alpha
_{S}))}{\xi (Ext(\alpha _{S}))}$,
\end{center}

\noindent which shows that the values of the $\mu $-contextual conditional
probability do not always depend on $\mu $-contexts.\medskip

Let us observe now that the above introduction of a probability measure on $%
L $ is purely formal. However, it can be intuitively justified by resorting
to the picture of the world provided in Section 3 when dealing with QM.
Indeed, whenever states are interpreted as equivalence classes of
preparation procedures, one can generalize the aforesaid picture and assume
that activating a preparation procedure $\pi $ produces an individual object
that is in a given state $S$ and, for every $\mu $-context $c$, possesses a
given set of properties depending on $c$, thus determining a truth
assignment $w$ on $L$. Activating again $\pi $\ produces another individual
object that still is in the state $S$, but may possess a different set of
properties for some $c\in \mathcal{C}$ (indeed we cannot control the
preparation procedure at a microscopic level, see (ii) in Section 3), thus
determining a truth assignment $w^{\prime }$ on $L$ that may differ from $w$%
. Given a universe $\mathcal{U}$ of individual objects, we can then maintain
that each individual object can be associated with a truth assignment on $L$
and that this correspondence is, generally, many-to-one. Let us roughly
reason in finite terms (we are only looking for an intuitive justification
of our mathematical structure here) and let us consider the set $Ext(\alpha
_{S})$ of all truth assignments that assign the value $t$ (\textit{true}) to
the atomic proposition $\alpha _{S}$ stating that the entity $H$ is in the
state $S$ (see Section 4). Then, we can assign a weight to $\alpha _{S}$\
that is proportional to the number of individual objects that are associated
with truth assignments in $Ext(\alpha _{S})$. Furthermore, similar
procedures lead to assign a weight to the atomic proposition $\alpha _{Ec}$\
stating that the entity $H$ has the property $E$ in the $\mu $-context $c$.
Hence a weight can be assigned to all propositions of $L$ following obvious
rules. It is thus apparent that the Definitions 5.1 and 5.2 formalize this
idea.

Whenever the above intuitive justification of the $\mu $-contextual
probability structure on $L$ is accepted, such structure can be seen as a
theoretical expression of the source of randomness described in Section 3,
(i), and it is important to stress that it is basically classical. Hence $%
\mu $-contextual conditional probabilities admit an epistemic
interpretation. In other words, they can be considered as indexes of our
lack of knowledge of the truth assignments on $\Pi $. In the framework of a
theory $\mathcal{T}\in \mathbb{T}$ characterized by the pair $(F,I)$ (see
Section 2), it may occur that these probabilities can be evaluated by using
the laws of $\mathcal{T}$. But, generally, they cannot be tested. Indeed,
one cannot know the physical situation at a microscopic level, hence the $%
\mu $-context associated, via $I$, with it. Therefore, $\mu $-contextual
probabilities must be considered as theoretical entities that can be
empirically interpreted only indirectly (see again Section 2). The next
Section is then devoted to discuss this issue in greater detail.

\section{Measurement procedures}

The predictions of a fully developed scientific theory are usually checked
by means of measurements whose theoretical description is part of the
theoretical language of the theory. In the present paper we are interested
in\ the theories in which $\mu $-contexts and tests of probabilities are
introduced. Hence, the formal apparatus of each theory of this kind must
include not only a $\mu $-contextual probability structure on $L$, but also
a theoretical description of the measurements that correspond to tests of
probabilities via the empirical interpretation of the theory (see Section
2). The physical preliminaries in Section 3 then provide us again with some
important suggestions. Firstly, a measurement\ may refer to more than one
atomic proposition simultaneously. Secondly, the theoretical description
must consider a subset of $\mu $-contexts that correspond to the possible
microscopic empirical situations underlying the test of probability.
Thirdly, a probability measure must be defined on the foregoing subset of $%
\mu $-contexts to take into account our limited knowledge of the microscopic
empirical situation when a test of probability is performed.

Bearing in mind the requirements above, we introduce the definition that
follows.\medskip

\textbf{Definition 6.1.} \textit{We denote by }$\mathbb{T}^{\mu \mathcal{M}}$%
\textit{\ the subclass of }$\mathbb{T}^{\mu }$\textit{\ characterized by the
following conditions.}

\textit{(i) For every }$\mathcal{T}\in \mathbb{T}^{\mu \mathcal{M}}$\textit{%
, a }$\mu $-\textit{contextual probability structure on }$L$\textit{\ is
defined.}

\textit{(ii) For every }$\mathcal{T}\in \mathbb{T}^{\mu \mathcal{M}}$\textit{%
, every }$E\in \mathcal{E}$\textit{\ is associated with a set }$\mathcal{M}%
_{E}$\textit{\ of }measurement procedures\textit{, and every }$M\in \mathcal{%
M}_{E}$ \textit{determines a\ }macroscopic measurement context $C_{M}$%
\textit{\ associated with a classical probability space }$(\mathcal{C}%
_{M},\Sigma _{M},\nu _{M})$\textit{, where }$\mathcal{C}_{M}$\textit{\ is a
subset of }$\mu $\textit{-contexts such that, for every }$c\in \mathcal{C}%
_{M}$\textit{, }$\left\{ c\right\} $ \textit{belongs to }$\Sigma _{M}$.%
\footnote{%
The theoretical notion of measurement procedure introduced here is rather
abstract because we want to avoid any reference to individual objects (see
Section 1). However, every quantum measurement of the kind considered in
Section 3 can be considered a measurement procedure in the sense defined
above.}

\textit{(iii) For every }$S\in \mathcal{S}$,\textit{\ }$\alpha _{S}\in \Pi
^{+}$.\textit{\medskip }

Generally, however, a test of probability refers to non-atomic (i.e.
molecular) propositions, which may require considering several atomic
propositions (hence several states, properties and $\mu $-contexts)
simultaneously, consistently with the first suggestion above. We are thus
naturally led to introduce the notions of \textit{compatibility}, \textit{%
testability} and \textit{joint testability} as follows.\medskip

\textbf{Definition 6.2.} \textit{Let us consider a theory }$\mathcal{T}\in 
\mathbb{T}^{\mu \mathcal{M}}$\textit{\ and a non-empty countable set }$%
\left\{ E,F,...\right\} \in 
%TCIMACRO{\U{2119} }%
%BeginExpansion
\mathbb{P}
%EndExpansion
(\mathcal{E})$\textit{\ of properties of }$L$\textit{. We say that }$E,F,...$%
\textit{\ are }compatible\textit{\ (in }$\mathcal{T}$\textit{) iff }$%
\mathcal{M}_{E}\cap \mathcal{M}_{F}\cap ...\neq \emptyset $\textit{.}

\textit{Moreover, for every }$A\in \Pi $\textit{, let }$\mathcal{E}%
_{A}=\left\{ E,F,...\right\} $\textit{\ be the (finite) set of all the
properties that occur (as indexes) in the formal expression of }$A$\textit{\
(together with indexes in }$\mathcal{C}$\textit{). We say that} $A$ \textit{%
is }testable\textit{\ (in }$\mathcal{T}$\textit{) iff the following
conditions hold.}

\textit{(i) No atomic state proposition occurs in }$A$\textit{\ (hence }$%
\mathcal{E}_{A}\neq \emptyset $).

\textit{(ii) }$E,F,...$\textit{\ are compatible.}

\textit{(iii) }$E,F,...$\textit{\ occur in the formal expression of A
together with the same index }$c$, \textit{and a measurement procedure }$%
M\in \mathcal{M}_{E}\cap \mathcal{M}_{F}\cap ...$ \textit{exists such that }$%
c\in \mathcal{C}_{M}$.

\textit{Then, we denote by }$\Pi _{\tau }$ \textit{the set of all testable
propositions of }$\Pi $, \textit{and for every }$A\in \Pi _{\tau }$\textit{\
we write }$A(c)$\textit{\ in place of }$A$\textit{\ whenever explicit
reference to the }$\mu $\textit{-context }$c$\textit{\ defined in (iii) must
be done.}

\textit{Finally, let }$\left\{ A,B,...\right\} $\ \textit{be a non-empty
finite set of propositions of }$\Pi _{\tau }$\textit{. We say that }$A,B,...$%
\textit{\ are }jointly testable\textit{\ (in }$\mathcal{T}$)\textit{\ iff
the proposition }$A\wedge B\wedge ...$\ \textit{is testable.\medskip }

Based on Definition 6.2 we state the following proposition.\medskip

\textbf{Proposition 6.1. }\textit{Let }$\mathcal{T}\in \mathbb{T}^{\mu 
\mathcal{M}}$. Then, the following statements hold in $\mathcal{T}$.

\textit{(i) Let us denote by }$k$\textit{\ the binary }compatibility relation%
\textit{\ on }$\mathcal{E}$\textit{\ defined by setting}

\begin{center}
\textit{\ for every }$E,F\in \mathcal{E}$\textit{, }$EkF$\textit{\ }iff$\ E$%
\textit{\ and }$F$ \textit{are compatible.}
\end{center}

\textit{Then, }$k$\textit{\ is reflexive and symmetric, but, generally, not
transitive.}

\textit{(ii) Let }$E\in \mathcal{E}$, $M\in \mathcal{M}_{E}$\textit{\ and }$%
c\in \mathcal{C}_{M}$.\textit{\ Then, the atomic proposition }$\alpha _{Ec}$%
\textit{\ belongs to }$\Pi _{\mathcal{\tau }}$.

\textit{(iii) Let }$A\in \Pi _{\tau }$\textit{, }$\mathcal{E}_{A}=\left\{
E,F,...\right\} $, $M\in \mathcal{M}_{E}\cap \mathcal{M}_{F}\cap ...$, $%
c_{0}\in \mathcal{C}_{M}$ \textit{and} $A=A(c_{0})$. \textit{Then, }$%
\mathcal{A}=\left\{ A(c)\mid c\in \mathcal{C}_{M}\right\} \subset \Pi _{\tau
}$ \textit{(equivalently, for every }$c\in \mathcal{C}_{M}$, $A(c)\in \Pi
_{\tau }$\textit{).\smallskip }

Proof. Straightforward.\textit{\medskip }

It remains to understand what one actually checks by means of a test of
probability performed by means of a measurement procedure $M\in \mathcal{M}%
_{E}\cap \mathcal{M}_{F}\cap ...$(to be precise, by means of the empirical
measurement procedure corresponding to $M$ via an empirical interpretation).
Therefore, let us resort again to the intuitive picture sketched in Section
3 with reference to QM. Such a picture suggests that, if a measurement is
performed of the (compatible) properties $E,F,...$\textit{\ }that occur in a
proposition $A(c_{0})\in \Pi _{\tau }$ by means of a measurement procedure $%
M\in \mathcal{M}_{E}\cap \mathcal{M}_{F}\cap ...$, then a $\mu $-context
occurs which one cannot control. Hence one cannot know whether the
measurement yields the truth value of $A(c_{0})$ or the truth value of
another proposition $A(c)\in \mathcal{A}$. When the measurement is iterated,
we obtain frequencies that approach a mean over $\mathcal{A}$ (hence over $%
\mathcal{C}_{M}$), in the large number limit, of the $\mu $-contextual
conditional probabilities defined in Section 5.

We add that we are generally interested in a class of theories in which all
tests corresponding to measurement procedures that belong to $\mathcal{M}%
_{E}\cap \mathcal{M}_{F}\cap ...$ yield the same results, which requires
that such procedures be probabilistically equivalent (e.g., the registering
devices considered in Section 3).

The following definition formalizes the above ideas.\medskip

\textbf{Definition 6.3.} \textit{Let }$\mathcal{T}\in \mathbb{T}^{\mu 
\mathcal{M}}$\textit{, let }$\Pi _{\mathcal{S}}=\left\{ A\in \Pi \mid 
\mathcal{E}_{A}=\emptyset \right\} $\textit{\ be the set of all propositions
in which no symbol of property occurs (briefly,} state-propositions\textit{%
), and let }$A$, $B\in \Pi _{\mathcal{\tau }}\cup \Pi _{\mathcal{S}}$\textit{%
.\ Then we introduce the following averages of }$\mu $\textit{-contextual
probabilities.}

\textit{(i) Let }$A$, $B$ \textit{be jointly testable, }$\mathcal{E}_{A}\cup 
\mathcal{E}_{B}=\left\{ E,F,...\right\} $, $M\in \mathcal{M}_{E}\cap 
\mathcal{M}_{F}\cap ...$, $c_{0}\in \mathcal{C}_{M}$ \textit{and} $%
A=A(c_{0}) $, $B=B(c_{0})$\textit{. Moreover, for every }$c\in \mathcal{C}%
_{M}$\textit{, let }$B(c)\in \Pi ^{+}$\textit{. We set}

\begin{center}
$<p(A\mid B)>_{\mathcal{C}_{M}}=\sum_{c\in \mathcal{C}_{M}}\nu
_{M}(\{c\})p(A(c)\mid B(c))$.
\end{center}

\textit{(ii) Let }$A\in \Pi _{\mathcal{\tau }}$\textit{, }$B\in \Pi _{%
\mathcal{S}}$, $\mathcal{E}_{A}=\left\{ E,F,...\right\} $, $M\in \mathcal{M}%
_{E}\cap \mathcal{M}_{F}\cap ...$, $c_{0}\in \mathcal{C}_{M}$ \textit{and} $%
A=A(c_{0})$. \textit{Moreover, let} $B\in \Pi ^{+}$\textit{. We set}

\begin{center}
$<p(A\mid B)>_{\mathcal{C}_{M}}=\sum_{c\in \mathcal{C}_{M}}\nu
_{M}(\{c\})p(A(c)\mid B)$.
\end{center}

\textit{(iii) Let }$A\in \Pi _{\mathcal{S}}$\textit{, }$B\in \Pi _{\mathcal{%
\tau }}$,\textit{\ }$\mathcal{E}_{B}=\left\{ E,F,...\right\} $, $M\in 
\mathcal{M}_{E}\cap \mathcal{M}_{F}\cap ...$, $c_{0}\in \mathcal{C}_{M}$ 
\textit{and} $B=B(c_{0})$. Moreover, let $B(c)\in \Pi ^{+}$ \textit{for
every }$c\in \mathcal{C}_{M}$\textit{. We set}

\begin{center}
$<p(A\mid B)>_{\mathcal{C}_{M}}=\sum_{c\in \mathcal{C}_{M}}\nu
_{M}(\{c\})p(A\mid B(c))$.
\end{center}

\textit{(iv) Let }$A,B\in \Pi _{\mathcal{S}}$\textit{\ and }$B\in \Pi ^{+}$%
\textit{. For every measurement procedure }$M$\textit{\ we set}

\begin{center}
$<p(A\mid B)>_{\mathcal{C}_{M}}=\sum_{cc\mathcal{C}_{M}}\nu
_{M}(\{c\})p(A\mid B)=p(A\mid B)$.
\end{center}

\textit{In case (iv) the reference to }$\mathcal{C}_{M}$\textit{\ in }$%
<p(A\mid B)>_{\mathcal{C}_{M}}$\textit{can be dropped. Moreover, we denote
by }$\mathbb{T}^{\mu \mathcal{MP}}$\textit{\ the subclass of }$\mathbb{T}%
^{\mu \mathcal{M}}$\textit{\ which consists of all theories satisfying the
condition that, if }$A,B\in \Pi _{\tau }\cup \Pi _{\mathcal{S}}$\textit{\
are such that }$\mathcal{E}_{A}\cup \mathcal{E}_{B}\neq \emptyset $\textit{,
and }$<p(A\mid B)>_{\mathcal{C}_{M}}$ \textit{is defined, then for every }$%
N\in M_{E}\cap M_{F}\cap ...$\textit{\ the average }$<p(A\mid B)>_{\mathcal{C%
}_{N}}$\textit{\ is also defined and coincides with }$<p(A\mid B)>_{\mathcal{%
C}_{M}}$. \textit{Therefore, whenever }$\mathcal{T}\in \mathbb{T}^{\mu 
\mathcal{MP}}$\textit{\ we drop the reference to }$\mathcal{C}_{M}$\textit{\
in }$<p(A\mid B)>_{\mathcal{C}_{M}}$\textit{, say that }$<p(A\mid B)>$%
\textit{\ is the }mean conditional probability\textit{\ of }$A$\textit{\
given }$B$\textit{\ and briefly write }$<p(A)>$\textit{\ in place of }$%
<p(A\mid B)>$\textit{\ if }$Ext(B)=W$.\medskip

By referring to Definition 6.3 we can maintain that, for every $\mathcal{T}%
\in \mathbb{T}^{\mu \mathcal{MP}}$, the empirical interpretation makes $M$
correspond to a \textit{mean probability test} that produces an outcome
which is expected to coincide with $<p(A\mid B)>$ in the large number limit.

We stress again that mean conditional probabilities are introduced in a
Kolmogorovian framework to take into account two different kinds of
ignorance. First, the lack of knowledge about the truth assignments on $\Pi $
mentioned at the end of Section 5. Second, the ignorance of the $\mu $%
-contexts to be associated with a probability test. Hence mean conditional
probabilities admit an \textit{epistemic} interpretation. Notwithstanding
this, they are not bound to satisfy Kolmogorov's assumptions, for they are
average quantities. In particular, it follows from Definition 6.3 that $%
<p(B)><p(A\mid B)>$ is generally different from $<p(A)><p(B\mid A)>$. Hence
a formal analogous of the Bayes theorem does not hold in the case of mean
conditional probabilities.

Finally, let us observe that the above definition of mean conditional
probabilities and mean probability tests are conceptually similar to the
universal averages and the universal measurements, respectively, introduced
by Aerts and Sassoli de Bianchi (2014, 2017). Moreover the recognition that
two kinds of lack of knowledge occur when a measurement is performed also
recalls the perspective proposed by these authors. As we have anticipated in
Section 1, we therefore make a brief comparison of our approach with Aerts
and Sassoli de Bianchi's in the Appendix.

\section{Quantum-like probability measures}

The set $\mathcal{E}$ of all properties is fundamental in every $\mathcal{T}%
\in \mathbb{T}^{\mu \mathcal{MP}}$. We intend to focus on it in the present
section and show that the notions and definitions in Section 6 allow us to
define, whenever some conditions on mean conditional probabilities are
satisfied, a family of quantum-like probability measures on $\mathcal{E}$
parametrized by the set of all states. To reach our aim, let us
preliminarily recall that, for all $E\in \mathcal{E}$, $M\in \mathcal{M}_{E}$
and $c\in \mathcal{C}_{M}$, $\alpha _{Ec}$ belongs to $\Pi _{\mathcal{\tau }%
} $ because of Proposition 6.1, (ii). Then we introduce the following
definition.\medskip

\textbf{Definition 7.1}. \textit{Let us consider a theory }$\mathcal{T}\in 
\mathbb{T}^{\mu \mathcal{MP}}$\textit{, let }$E\in \mathcal{E}$\textit{, }$%
M\in \mathcal{M}_{E}$\textit{, }$c\in \mathcal{C}_{M}$\textit{, }$S\in 
\mathcal{S}$\textit{,} \textit{and let }$P_{S}(E)$\textit{\ be the mean
conditional probability of }$\alpha _{Ec}$\textit{\ given }$\alpha _{S}$%
\textit{, that is,}

\begin{center}
$P_{S}(E)=<p(\alpha _{Ec}\mid \alpha _{S})>=\sum_{c\in \mathcal{C}_{M}}\nu
_{M}(\{c\})p(\alpha _{Ec}\mid \alpha _{S})\medskip $

$=\sum_{c\in \mathcal{C}_{M}}\nu _{M}(\{c\})\frac{\xi (Ext(\alpha _{Ec})\cap
Ext(\alpha _{S}))}{\xi (Ext(\alpha _{S}))}$\textit{.}
\end{center}

\textit{Then, \ we denote by }$\prec $\textit{\ and }$\approx $\textit{\ the
preorder and the equivalence relation on }$\mathcal{E}$\textit{,
respectively, defined by setting, for every }$E,F\in \mathcal{E}$\textit{,}

\begin{center}
$E\prec F$\textit{\ }iff\textit{, for every }$S\in \mathcal{S}$\textit{, }$%
P_{S}(E)\leq P_{S}(F)$
\end{center}

\noindent \textit{and}

\begin{center}
$E\approx F$\textit{\ }iff\textit{\ }$E\prec F$\textit{\ and }$F\prec E$%
\textit{.\medskip }
\end{center}

It is now important to consider a special case that allows us to place
physical theories as CM, SM and QM within the general framework constructed
in Sections 4-6. To this end we introduce the following definition.\medskip

\textbf{Definition 7.2}. \textit{Let us consider a theory }$\mathcal{T}\in 
\mathbb{T}^{\mu \mathcal{MP}}$\textit{\ such that }$\prec $\textit{\ is a
partial order and }$(\mathcal{E},\prec )$\textit{\ is an orthocomplemented
lattice. We denote }meet\textit{, }join\textit{, }orthocomplementation%
\textit{, }least element\textit{\ and }greatest element\textit{\ of }$(%
\mathcal{E},\prec )$\ \textit{by }$\Cap $\textit{, }$\Cup $\textit{, }$%
^{\bot }$\textit{, }$\mathsf{O}$\textit{\ and }$\mathsf{U}$\textit{,
respectively. Moreover, we denote by }$\bot $\textit{\ the (binary)
orthogonality relation canonically induced by }$^{\bot }$\textit{\ on }$(%
\mathcal{E},\Cap \mathit{,}\Cup \mathit{,}^{\bot })$\textit{\footnote{%
We recall that $^{\bot }\mathcal{\ }$is a unary operation on $(\mathcal{E}%
,\prec )$ such that, for every $E,F\in \mathcal{E}$, $E^{\bot \bot }=E$, $%
E\prec F$ implies $F^{\bot }\prec E^{\bot }$, $E\Cap E^{\bot }=\mathsf{O}%
\mathit{\ }$and $E\Cup E^{\bot }=\mathsf{U}$. Then $\bot $\ is the
non-reflexive and symmetric binary relation on $\mathcal{E}$\ defined by
setting, for every $E,F\in \mathcal{E}$, $E\bot F$ iff $E\prec F^{\bot }.$}.
Then, for every }$S\in \mathcal{S}$\textit{, we say that the mapping}

\begin{center}
$P_{S}:\mathcal{E}\longrightarrow \lbrack 0,1],E\longrightarrow
P_{S}(E)=<p(\alpha _{Ec}\mid \alpha _{S})>$
\end{center}

\noindent \textit{is a }generalized probability measure\textit{\ on }$(%
\mathcal{E},\Cap \mathit{,}\Cup \mathit{,}^{\bot })$\ \textit{iff it
satisfies the following conditions.}

\textit{(i) }$P_{S}(\mathsf{U})=1$.

\textit{(ii) If }$\left\{ E_{i}\right\} _{i\in N}$\textit{\ is a family of
properties that are pairwise disjoint (i.e., for every }$k,l\in N$\textit{, }%
$E_{k}\bot E_{l}$),\textit{\ then}

\begin{center}
$P_{S}(\Cup _{i}E_{i})=\sum_{i}P_{S}(E_{i})$.
\end{center}

\textit{Let }$E\in \mathcal{E}$. \textit{Whenever }$P_{S}$\textit{\ is a
generalized probability measure on }$(\mathcal{E},\Cap \mathit{,}\Cup 
\mathit{,}^{\bot })$, \textit{we say that }$P_{S}(E)$\textit{\ is the }%
q-probability\textit{\ of }$E$\textit{\ given}$\mathcal{\ }S$\textit{%
.\medskip }

Definition 7.2 implies that a generalized probability measure $P_{S}$\ is a
classical probability measure only if $(\mathcal{E},\Cap \mathit{,}\Cup 
\mathit{,}^{\bot })$\ is a Boolean lattice. Hence, generally, $P_{S}$\ does
not satisfy Kolmogorov's assumptions. Nevertheless, the q-probability $%
P_{S}(E)$\ of a property $E\in \mathcal{E}$\ given $S$ admits an epistemic
interpretation and can be empirically tested, as it is a special case of the
mean conditional probability introduced in Definition 6.3. It is then
natural to wonder whether a \textit{conditional q-probability} of a property 
$E\in \mathcal{E}$ given another property $F\in \mathcal{E}$\ can be defined
by means of $P_{S}$, generalizing standard procedures in classical
propositional logic. But if one tries to put

\begin{center}
$P_{S}(E\mid F)=\frac{P_{S}(E\Cap F)}{P_{S}(F)}$,
\end{center}

\noindent then the mapping

\begin{center}
$P_{SF}:E\in \mathcal{E}\longrightarrow P_{S}(E\mid F)\in \lbrack 0,1]$
\end{center}

\noindent is not a generalized probability measure on $(\mathcal{E},\Cap 
\mathit{,}\Cup \mathit{,}^{\bot })$\ whenever this lattice is not boolean.
Indeed, consider a property $E=E_{1}\Cup E_{2}$, with $E_{1},E_{2}\in 
\mathcal{E}$ and $E_{1}\bot E_{2}$. We obtain

\begin{center}
$P_{SF}(E)=P_{SF}(E_{1}\Cup E_{2})=P_{S}(E_{1}\Cup E_{2}\mid F)=\frac{%
P_{S}((E_{1}\Cup E_{2})\Cap F)}{P_{S}(F)}$,
\end{center}

\noindent which is generally different from

\begin{center}
$\frac{P_{S}((E_{1}\Cap F)\Cup (E_{2}\Cap F))}{P_{S}(F)}=P_{S}(E_{1}\mid
F)+P_{S}(E_{2}\mid F)=P_{SF}(E_{1})+P_{SF}(E_{2})$
\end{center}

\noindent whenever $(\mathcal{E},\Cap \mathit{,}\Cup \mathit{,}^{\bot })$\
is not distributive.\medskip

One can, however, introduce a non-standard kind of conditional probability
by considering mean probability tests performed in sequence rather than
conjointly. Indeed one can again draw inspiration from QM and single out
theories in $\mathbb{T}^{\mu \mathcal{MP}}$ where measurement procedures
exist which correspond, via empirical interpretation, to mean probability
tests that filter the sample of the entity that is considered in a prefixed
way, producing a new sample on which the same or a different test can be
performed. Moreover, still inspired by QM, we are interested in those mean
probability tests that yield frequency 1 when repeated on the new sample. We
therefore introduce the following definition.\medskip

\textbf{Definition 7.3}. \textit{Let us consider a theory }$\mathcal{T}\in 
\mathbb{T}^{\mu \mathcal{MP}}$\textit{\ and for every }$F\in \mathcal{E}$ 
\textit{let us put }$\mathcal{S}_{F}=\left\{ S\in \mathcal{S}\mid
P_{S}(F)\neq 0\right\} $\textit{. Then we say that a measurement procedure} $%
M\in \mathcal{M}_{F}$\ \textit{is of }first kind\textit{\ iff it is
associated with a mapping}

\begin{center}
$t_{F}:\mathcal{S}_{F}\longrightarrow \mathcal{S}_{F},S\longrightarrow
t_{F}(S)$
\end{center}

\noindent \textit{such that }$P_{t_{F}(S)}(F)=1$. For every $E\in \mathcal{E}
$\textit{\ and first kind measurement procedure }$M\in \mathcal{M}_{F}$%
\textit{\ we then put }

\begin{center}
$P_{S}(E\Vert F)=P_{t_{F}(S)}(E)$\textit{.}
\end{center}

\textit{Moreover, let }$(\mathcal{E},\prec )$\textit{\ be an
orhocomplemented lattice in }$\mathcal{T}$\textit{\ and let }$P_{S}$\textit{%
\ and }$P_{t_{F}(S)}$\textit{\ be generalized probability measures on }$(%
\mathcal{E},\prec )$\textit{. Then we say that }$P_{S}(E\Vert F)$\textit{\
is the }conditional q-probability\textit{\ of }$E$\textit{\ given }$F$%
\textit{\ and }$S$\textit{.\medskip }

If a theory $\mathcal{T}\in \mathbb{T}^{\mu \mathcal{MP}}$ contains a first
kind measurement procedure $M\in \mathcal{M}_{F}$, $S\in \mathcal{S}_{F}$
and $P_{S}(E\Vert F)$ is defined, then $P_{S}(E\Vert F)$ can be tested, as
mean probability tests always exist for $P_{t_{F}(S)}(E)$ (see Section 6;
however, no analogous of the Bayes theorem can be stated for conditional
q-probabilities). Definition 7.3 thus introduces a non-standard conditional
probability on $(\mathcal{E},\prec )$ that can be tested and coexists with
the $\mu $-contextual conditional probability introduced in Definition 5.2,
which instead cannot be tested directly and has the status of a purely
theoretical notion.

\section{Physical theories}

The mathematical apparatus worked out in the previous sections has been
contrived bearing in mind CM, SM and QM. Indeed, the language of these
theories contains terms denoting entities, properties, states and
measurements, hence CM, SM and QM belong to $\mathbb{T}$ (see Definition
4.1). Moreover, microscopic contexts can be introduced in CM, SM and QM as
special cases of physical systems, hence these theories belong to $\mathbb{T}%
^{\mu }$ (ib.). By referring then to Definition 6.1, we see that conditions
(i), (ii) and (iii) characterizing $\mathbb{T}^{\mu \mathcal{M}}$ are
compatible with (even if not implied by) CM, SM and QM (condition (iii), in
particular, establishes that no "useless" state, i.e., a state $S$ such that 
$p(\alpha _{S})=0$, occurs in a theory $\mathcal{T}\in \mathbb{T}^{\mu 
\mathcal{M}}$). Finally, the condition characterizing $\mathbb{T}^{\mu 
\mathcal{MP}}$\ in Definition 6.3 states that, for every $\mathcal{T}\in 
\mathbb{T}^{\mu \mathcal{MP}}$, the mean conditional probability of a
proposition $A$ given a proposition $B$ does not depend on the choice of the
measurement procedure, which also is compatible with CM, SM and QM.

Based on the above arguments, we can now assume that CM, SM and QM belong to 
$\mathbb{T}^{\mu \mathcal{MP}}$. It is then easy to see that all the notions
introduced in the previous sections collapse into standard notions in the
case of CM and SM. In the case of QM, instead, our assumption explains some
relevnt aspects of this theory and of quantum probability in terms of the
general notions introduced in $\mathbb{T}^{\mu \mathcal{MP}}$. We therefore
discuss these issues in the following sections, referring to CM and QM only
and leaving apart SM, which can be dealt with by extending our treatment of
CM in an obvious way.

\section{Classical mechanics}

Let us begin by listing some basic features of CM, some of which can be
deduced at once from the phase space representation of states and properties.

(i) CM deals with individual objects, their (pure) states and (physical)
properties. Both macroscopic and microscopic measurement contexts can be
introduced in it and supplied with an intuitive (intended) interpretation,
but it is assumed that each individual object either possesses or does not
possess any property that is considered, independently of any measurement
procedure.

(ii) Whenever the state $S$ of an individual object $x$ is given, the set of
all properties possessed by $x$ is determined by $S$, and it is different
from the set of properties possessed by another individual object in a state 
$S^{\prime }$ different from $S$.

(iii) For every finite set $\left\{ E,F,...\right\} $ of properties and
individual object $x$, one can check (at least in principle) which
properties in $\left\{ E,F,...\right\} $\ are possessed by $x$ and which are
not by performing an (exact) measurement that consists in measuring
simultaneously $E,F,...$ .

(iv) Different properties can be assumed to have different phase space
representations, hence they are not equivalent, in the sense that\ there are
individual objects that possess one of them and not the other.

(v) Every negation of a proposition stating a property is a proposition
stating a property, and every (finite) conjunction or disjunction of
propositions stating properties is a proposition stating a property (see,
e.g., Garola and Sozzo, 2013).

(vi) For every property $E$ and individual object $x$, a measurement exists
(at least as a limit of real measurements) which establishes whether $x$
possesses or does not possess the property $E$ without perturbing the state $%
S$ of $x$.

Let us discuss now how the general notions introduced in Sections 4-7
specialize in the case of CM.

First of all, (i) implies that macroscopic and microscopic contexts play no
role in the truth assignments on $\Pi $. Hence,\ for every $w\in W$, $E\in 
\mathcal{E}$\textit{, }$c,d\in \mathcal{C}$, the equality $w(\alpha
_{Ec})=w(\alpha _{Ed})$ holds in CM, which implies $\alpha _{Ec}\equiv
\alpha _{Ed}$ (see Definition 4.2), $Ext(\alpha _{Ec})=Ext(\alpha _{Ed})$
(see Definition 5.1) and $\xi (Ext(\alpha _{Ec}))=\xi (Ext(\alpha _{Ed}))$
(see Definition 5.2). Therefore, we can drop any reference to $\mu $%
-contexts in the following. In particular, we write $\alpha _{E}$ and $\Pi _{%
\mathcal{E}}^{a}$\ in place of $\alpha _{EC}$ and $\Pi _{\mathcal{EC}}^{a}$,
respectively, and notice that the mapping $\tau :\mathcal{E}\longrightarrow
\Pi _{\mathcal{E}}^{a},E\longrightarrow \alpha _{E}$ is bijective.

Secondly, (ii) implies that, for every $w\in W$ and $S\in \mathcal{S}$%
\textit{,} the requirement $w(\alpha _{S})=t$ determines the values of $w$
on all (atomic and molecular) propositions of $\Pi $: in particular, $%
w(\alpha _{S^{\prime }})=f$ for every $S^{\prime }\neq S$. Hence $Ext(\alpha
_{S})$ (see Definition 5.1) is a singleton, whose unique element we denote
by $w_{S}$. It follows from Definition 5.2 that, for every $E\in \mathcal{E}$%
\textit{, }$p(\alpha _{E}\mid \alpha _{S})\in \left\{ 0,1\right\} $.
Moreover, every truth assignment on $\Pi $\ refers to individual objects in
a state $S$, hence for every $w\in W$ a state $S\in \mathcal{S}$ exists such
that $w=w_{S}$. Therefore, the mapping $s:\mathcal{S\longrightarrow }W,S%
\mathcal{\longrightarrow }w_{S}$ is bijective.

Thirdly, let us come to measurements. Then, (iii) implies that, for every
countable set $\left\{ E,F,...\right\} \in 
%TCIMACRO{\U{2119} }%
%BeginExpansion
\mathbb{P}
%EndExpansion
(\mathcal{E})$,\ the properties $E,F,...$ are compatible in the sense
established in Definition 6.2. Moreover, every proposition $A\in \Pi $, such
that no atomic state proposition occurs in it, is testable, that is, $A\in
\Pi _{\tau }$, and for every non-empty finite set $\left\{ A,B,...\right\}
\in 
%TCIMACRO{\U{2119} }%
%BeginExpansion
\mathbb{P}
%EndExpansion
(\Pi _{\tau })$, the propositions $A,B,...$ are jointly testable and the
result of an exact measurement neither depends on the macroscopic context
nor on the $\mu $-contexts. Therefore, for every $A,B\in \Pi _{\tau }\cup
\Pi _{\mathcal{S}}$\textit{,} if\textit{\ }$B\in \Pi ^{+}$ the mean
conditional probability $<p(A\mid B)>$ is defined (see Definition 6.3) and
coincides with $p(A\mid B)$. The notion of mean conditional probability thus
reduces to the notion of conditional probability. Moreover, the mapping $%
P_{S}$ introduced in Definition 7.2 is such that, for every $E\in \mathcal{E}
$ and $S\in \mathcal{S}$,

\begin{center}
$P_{S}(E)=p(\alpha _{E}\mid \alpha _{S})\in \left\{ 0,1\right\} $.
\end{center}

The results obtained above imply that, for every $E,F,...\in \mathcal{E}$, $%
E\prec F$\ (see Definition 7.1) iff $\alpha _{E}<\alpha _{F}$\ (see
Definition 4.3). Indeed, let us recall that we have assumed in Definition
6.1 that, for every $S\in \mathcal{S}$, $\alpha _{S}\in \Pi ^{+}$, hence $%
\xi (Ext(\alpha _{S}))\neq 0$, which implies $\xi (\left\{ w_{S}\right\}
)\neq 0$ in CM. Therefore, for every $E,F\in \mathcal{E}$, the following
sequence of coimplications holds.

\begin{center}
$(E\prec F)$ iff (for every $S\in \mathcal{S}$, $p(\alpha _{E}\mid \alpha
_{S})\leq p(\alpha _{F}\mid \alpha _{S})$) iff (for every $S\in \mathcal{S}$%
, $\xi (Ext(\alpha _{E})\cap Ext(\alpha _{S}))\leq \xi (Ext(\alpha _{F})\cap
Ext(\alpha _{S})$) iff (for every $w_{S}\in W$, $\xi (Ext(\alpha _{E})\cap
\left\{ w_{S}\right\} )\leq \xi (Ext(\alpha _{F})\cap \left\{ w_{S}\right\}
) $) iff ($Ext(\alpha _{E})\subset Ext(\alpha _{F})$) iff ($\alpha
_{E}<\alpha _{F}$).
\end{center}

It follows that the order structures $(\mathcal{E,}\prec )$\ and $(\Pi _{%
\mathcal{E}}^{a},<)$ are isomorphic. Moreover, $<$ and $\prec $\ are partial
orders. Indeed, (iv) implies that, for every $E,F\in \mathcal{E}$, if $E\neq
F$ then there is a truth assignment\ $w_{S}$\ which assigns the value $t$ to
one of the propositions $\alpha _{E}$ and $\alpha _{F}$, and value $f$ to
the other. Hence $\alpha _{E}\equiv \alpha _{F}$ iff $\alpha _{E}=\alpha
_{F} $, which implies that $<$\ is a partial order on $\Pi _{\mathcal{E}%
}^{a} $. Because of the aforesaid isomorphism, also $\prec $\ is a partial
order.

Let us consider now q-probability. Because of (v), $(\Pi _{\mathcal{E}%
}^{a},<)$ is a Boolean lattice, hence $(\mathcal{E,}\prec )$ is a Boolean
lattice (whose meet, join and complementation we denote now, by abuse of
language, by $\cap $, $\cup $ and $^{c}$, respectively). Thus, for every $%
S\in \mathcal{S}$, the q-probability $P_{S}$ is a classical probability
measure on $(\mathcal{E,}\cap ,\cup ,^{c})$ (the proof of this statement
follows at once from Proposition 5.2 because of the equality and isomorphism
above). Therefore, for every $E,F\in \mathcal{E}$, the conditional
probability in the state $S$ of $E$ given $F$ can be defined in a standard
way, as follows

\begin{center}
$P_{S}(E\mid F)=\frac{P_{S}(E\cap F)}{P_{S}(F)}\in \left\{ 0,1\right\} $
\end{center}

\noindent (where $P_{S}(F)\neq 0$, hence $P_{S}(F)=1$).

Finally, let us consider the conditional q-probability $P_{S}(E\Vert F)$.
Because of (vi), a first kind measurement procedure exists for every $F\in 
\mathcal{E}$ such that $t_{F}$ (see Definition 7.3) is the identity mapping.
We obtain in this case that $P_{S}(E\Vert F)=P_{t_{F}(S)}(E)=P_{S}(E)$.
Since $P_{S}(E)\in \left\{ 0,1\right\} $, it is easy to see that $%
P_{S}(E)=P_{S}(E\mid F)$. Notwithstanding this equality, however, there is a
the deep conceptual difference between standard conditional probability and
conditional q-probability.

\section{Quantum mechanics}

We have assumed in Section 8 that CM, SM and QM belong to $\mathbb{T}^{\mu 
\mathcal{MP}}$. Yet, at variance with CM and SM, QM is a theory in which
(macroscopic) contexts play a fundamental role. Whenever $L$ is assumed to
be the basic language of QM, contextuality implies that the inequality $%
w(\alpha _{Ec})\neq w(\alpha _{Ed})$ (which implies $Ext(\alpha _{Ec})\neq
Ext(\alpha _{Ed})$) holds in $L$ for some $w\in W$, $E\in \mathcal{E}$%
\textit{, }$c,d\in \mathcal{C}$.

Let us consider now Hibert space QM (HSQM). Within HSQM each entity
(physical system) is associated with a Hilbert space $\mathcal{H}$, each
state $S$ is represented by a density operator $\rho _{S}$ on $\mathcal{H}$
and each property $E$ is represented by an orthogonal projection operator $%
P_{E}$ on $\mathcal{H}$. Since the set of all orthogonal projection
operators on $\mathcal{H}$ is an orthomodular lattice in which a partial
order is defined independently of any probability measure, this
representation induces on $\mathcal{E}$\ an order, that we denote by $\ll $,
and $(\mathcal{E,}\ll )$\ \ is an orthomodular lattice (the standard quantum
logic mentioned in Section 1). Moreover, Born's rule associates a
probability value $Tr\left[ \rho _{S}P_{E}\right] $ (which does not depend
on any context) with every pair $(E,S)$, hence a quantum probability

\begin{center}
$Q_{S}:\mathcal{E}\longrightarrow \lbrack 0,1],E\longrightarrow Tr\left[
\rho _{S}P_{E}\right] $
\end{center}

\noindent is defined which is said to be a \textit{generalized probability
measure} on $(\mathcal{E,}\ll )$\ (see, e.g., Beltrametti and Cassinelli,
1981), and the family $\left\{ Q_{S}\right\} _{S\in \mathcal{S}}$ is \textit{%
ordering} on $(\mathcal{E,}\ll )$ (ib.), which means that the order induced
by it on $\mathcal{E}$ coincides with $\ll $. Therefore, the lattice
structure of $(\mathcal{E,}\ll )$ can be seen as induced by $\left\{
Q_{S}\right\} _{S\in \mathcal{S}}$. This feature of HSQM implies that the
order $\ll $\ and the probability $Q_{S}$ can be considered as the specific
forms that the order $\prec $\ and the mapping $P_{S}$, respectively, take
in QM (see Definitions 7.1 and 7.2). We thus obtain in QM

\begin{center}
$P_{S}(E)=<p(\alpha _{EC}\mid \alpha _{S})>=Q_{S}(E)=Tr\left[ \rho _{S}P_{E}%
\right] $.
\end{center}

Furthermore, if the quantum probability $Q_{S}$ replaces $P_{S}$ in the
conditions (i) and (ii) stated in Definition 7.2, then these conditions are
satisfied, which makes the above classification of $Q_{S}$\ as generalized
probability measure consistent with Definition 7.2. We thus obtain an
interpretation of quantum probability measures that leads to consider them
mean conditional probabilities. They can therefore be seen as derived
notions within a Kolmogorovian framework, as we have anticipated in Section
1, which explains their non-classical character but shows that they admit an
epistemic interpretation, at variance with their standard ontic
interpretation (see Section 6).

In addition, let us denote by $\kappa $ the compatibility relation
introduced in QM on the set of all properties by setting, for every pair $%
(E,F)$ of properties, $E\kappa F$ iff $\left[ P_{E},P_{F}\right] =0$. This
relation is reflexive and symmetric but not transitive. Hence it can be
considered as the specific form that the relation $k$ introduced in
Proposition 6.1 takes in QM.

Coming to measurements, let us recall that first kind measurement procedures
exist in QM (see, e.g., Piron, 1976; Beltrametti and Cassinelli, 1981) and
that the L\"{u}ders rule states that, whenever an (ideal) first kind
measurement of a property $E$ is performed on an ensemble described by $\rho
_{S}$, the subensemble that passes the measurement is described by the
density operator $\frac{P_{E}\rho _{S}P_{E}}{Tr\left[ \rho _{S}P_{E}\right] }
$. Let us therefore denote by $D\left( \mathcal{H}\right) $ the set of all
density operators on $\mathcal{H}$. Then the mapping

\begin{center}
$\tau _{E}:D\left( \mathcal{H}\right) \longrightarrow D\left( \mathcal{H}%
\right) ,\rho _{S}\longrightarrow \frac{P_{E}\rho _{S}P_{E}}{Tr\left[ \rho
_{S}P_{E}\right] }$
\end{center}

\noindent can be considered as the specific form that the mapping $t_{E}$
introduced in Definition 7.3 takes in QM.

Finally, we recall that the conditional probability $Q_{S}(F\mid E)$, in a
state $S$, of a property $F$ given a property $E$, is defined in QM by
referring to a measurement of $F$ after a measurement of $E$ on an ensemble
described by $\rho _{S}$, and is given by $\frac{Tr\left[ P_{F}P_{E}\rho
_{S}P_{E}P_{F}\right] }{Tr\left[ P_{E}\rho _{S}P_{E}\right] }$. Hence this
quantity can be considered as the specific form that the conditional
q-probability of $F$ given $E$ and $S$ introduced in Definition 7.3 takes in
QM. We thus obtain

\begin{center}
$P_{S}(F\Vert E)=Q_{S}(F\mid E)=\frac{Tr\left[ P_{F}P_{E}\rho _{S}P_{E}P_{F}%
\right] }{Tr\left[ P_{E}\rho _{S}P_{E}\right] }$.
\end{center}

\section{Conclusions}

According to the perspective presented in this paper, a class $\mathbb{T}%
^{\mu \mathcal{MP}}$ of scientific theories can be singled out in which mean
probabilities that do not satisfy the assumptions of Kolmogorov's
probability theory may occur within a Kolmogorovian probabilistic framework
because of contextuality. The conditions characterizing $\mathbb{T}^{\mu 
\mathcal{MP}}$ are compatible with CM, SM and QM, which therefore can be
maintained to belong to this class of theories. In the case of QM this
membership implies that quantum probability measures can be seen as mean
conditional probabilities that have a non-classical structure but admit an
epistemic interpretation, which challenges the standard ontic interpretation
of quantum probability. In addition, we also obtain that some typical
features of QM, as the compatibility relation on the set of all physical
properties and the quantum notion of conditional probability, are special
cases of general notions introduced in $\mathbb{T}^{\mu \mathcal{MP}}$.
These results are obtained without referring to individual objects, which
makes them hold even if only a minimal interpretation of QM is accepted to
avoid the problems of the standard quantum theory of measurement.\medskip

\textbf{Acknowledgements}. We thank Dr. Antonio Negro, Prof. Carlo Sempi and
Dr. Karin Verelst for valuable suggestions and discussions.\medskip

\begin{center}
APPENDIX\smallskip
\end{center}

As anticipated in Sections 1 and 6, we intend to make here a brief
comparison of our approach to quantum probability with Aerts and Sassoli de
Bianchi's solution to the measurement problem of QM. This solution was
expounded, in particular, in a technical paper (Aerts and Sassoli de
Bianchi, 2014) and in a book aiming to make it understandable to a wider
audience (Aerts and Sassoli de Bianchi, 2017). Our comparison will be made
mainly referring to that book, which will make our description of the
similarities and differences between the two approaches simple and intuitive.

To begin with, let us recall that the proposal of Aerts and Sassoli de
Bianchi finds its roots in Aerts' \textit{hidden measurements} idea (see,
e.g., Aerts, 1986). By developing this idea Aerts and Sassoli de Bianchi
provide a detailed description of the measurement process in QM by
constructing an elaborate model whose core is the Bloch representation of
the pure states of a spin $%
%TCIMACRO{\U{bd}}%
%BeginExpansion
{\frac12}%
%EndExpansion
$ physical system. This representation, in which every pure state
corresponds to a point on the surface of a three-dimensional sphere, is
extended by Aerts and Sassoli de Bianchi by considering the points within
the Bloch sphere as representative of new states, considered as pure rather
than mixed, as it would occur instead in the standard formalism of QM. The
final action of an instrument measuring the spin of the physical system is
then represented in the sphere by means of an elastic band connecting the
north with the south pole of the sphere. When the measurement is performed,
the state of the system moves orthogonally onto the elastic band and sticks
to it. Then, the elastic band breaks in a point whose position on the band
is unpredictable, leading the state either on the north pole (\textit{spin up%
}) or on the south pole (\textit{spin down}), depending on the position of
the breaking point.

To make quantitative the above qualitative description of the measurement
process, Aerts and Sassoli de Bianchi assume that the elastic band is
characterized by a probability distribution whose value in a given point of
the band is interpreted as the density of probability of breaking at that
point. Moreover, when the measurement is repeated, the properties of the new
elastic band may be different from the properties of the old one, and in
this case the new band is characterized by a different probability
distribution. Hence, when the measurement is repeated many times on spin $%
%TCIMACRO{\U{bd}}%
%BeginExpansion
{\frac12}%
%EndExpansion
$ systems in a given state, to predict the frequency of each possible
outcome one must average over all probability distributions. The authors
call this average \textit{universal average}, and then call the measurement 
\textit{universal measurement}. A quantum measurement of the spin of the
physical system is then assumed to be a measurement of this kind.

The following remarks are now important.

(i) The experimenter can choose to perform a measurement but cannot choose
the breaking point of the elastic band, as he takes every possible
precaution to avoid influencing the outcome. The breaking point of the band
is instead assumed by Aerts and Sassoli de Bianchi to be the result of
nondeterministic and unpredictable environmental fluctuations. Hence the
elastic band corresponds to a \textit{potentiality region} and physical
quantities do not pre-exist to the measurement but are \textit{actualized}
by it. Therefore the authors consider quantum probability as \textit{ontic},
which fits in well with the standard interpretation of QM.

(ii) In the description of the measuring process probability occurs twice.
Firstly, when the elastic band is characterized by a probability
distribution. Secondly, when averaging over probability distributions to
obtain a universal average, which intuitively means that every possible
distribution has the same probability to occur every time the measurement is
repeated (because of (i) the experimenter does not influence in any way the
\textquotedblleft emerging\textquotedblright\ of a specific distribution).

(iii) Aerts and Sassoli de Bianchi's description of spin $%
%TCIMACRO{\U{bd}}%
%BeginExpansion
{\frac12}%
%EndExpansion
$ measurements is not a hidden-variables theory in a standard sense. Indeed,
it explains the randomness of the observed outcomes as a consequence of
fluctuations in the measuring system, consistently with the Aerts' idea of
hidden measurements mentioned above, rather than a consequence of our
incomplete knowledge of the real state of the measured entity.

After constructing the above model, Aerts and Sassoli de Bianchi make a
considerable effort to generalize it to physical entities whose measurements
can give more than two possible outcomes. In this case the mathematical
apparatus becomes much more complicated (in particular, the Bloch sphere
becomes a hypersphere in a space with more than three dimensions and the
elastic band is substituted by a hypermembrane). Nevertheless the basic
features of spin $%
%TCIMACRO{\U{bd}}%
%BeginExpansion
{\frac12}%
%EndExpansion
$ measurements pointed out above remain unchanged, hence we will refer to
them in the following without entering the details of the general model.

Let us come to our approach. Here a canonical distinction between preparing
and registering devices is introduced (Section 3) but no explicit model for
the measurement process is proposed. Rather, a very simple picture assuming
the existence of a microscopic world underlying the macroscopic world of our
everyday experience is provided to intuitively justify our formalism
(Sections 3 and 5). According to this picture, we do not know what is going
on at a microscopic level neither in preparing nor in registering devices.
To deal with the first kind of lack of knowledge, we introduce a probability
measure on the language $L$ of the theories that we are considering, which
corresponds to the assignments of probability distributions on elastic bands
in Aerts and Sassoli de Bianchi's description. Yet, our probability measure
is introduced because the quantum description of the state of a physical
system is maintained to be incomplete, according to the spirit of standard
hidden variables theories (but only context-depending propositions occur in $%
L$, which implies that the \textquotedblleft no go\textquotedblright\
theorems mentioned in footnote 1 do not apply): hence, it is considered 
\textit{epistemic}. To deal with the second kind of lack of knowledge we
introduce $\mu $-contexts, which complies with the hidden measurements idea
and parallels the introduction of different elastic bands when the
measurment is repeated in Aerts and Sassoli de Bianchi's description. Mean
conditional probabilities then parallel universal averages. We, however, do
not introduce any assumption of equiprobability (see (ii) above), which
makes our approach slightly more general. More important, mean conditional
probabilities bear an epistemic interpretation, for they are classical
weighted means of epistemic probabilities, at variance with Aerts and
Sassoli de Bianchi universal averages.

To close, let us recall that Aerts also introduced \textit{state property
systems} (see Aerts, 1999 and related bibliography), which successively
evolved in the \textit{state-context-property} (SCoP) formalism (see, e.g.,
Aerts and Gabora, 2005; such a formalism was mainly used for working out a
theory of concepts, in particular in the field of quantum cognition). Then,
the SCoP formalism can be (partially) translated into the formalism
developed in the present paper, and conversely. Indeed, its basic structure
can be summarized as follows.

(i) Fundamental notions: \textit{entity},\textit{\ state}, \textit{%
(measurement)} \textit{context}, \textit{property} (hence the SCoP formalism
characterizes a class of theories that also belong to $\mathbb{T}$, see
Definition 4.1).

(ii) Fundamental definitions: set of states $\Sigma $, set of contexts $M$,
set of properties L; entity $(\Sigma ,M,L,\mu ,\nu )$, where $\mu :\Sigma
\times M\times \Sigma \longrightarrow \lbrack 0,1],(p,e,q)\longrightarrow
\mu (p,e,q)$ is a \textit{state-transition probability function} that
represents the likelihood to transition from the state $p$ to the state $q$
under the influence of the context $e$, and $\nu :\Sigma \times
L\longrightarrow \lbrack 0,1],(p,a)\longrightarrow \nu (p,a)$ is a \textit{%
property-applicability function} that estimates how applicable is the
property $a$ to the state $p$ of the entity.

Then, based on the physical interpretation of the SCoP formalism, the
following bijective correspondences with the formalism introduced in the
present paper can be established.

$c_{1}:\Sigma \longrightarrow \mathcal{S},p\longrightarrow S$,

$c_{2}:M\longrightarrow \cup _{E\in \mathcal{E}}\mathcal{M}%
_{E},e\longrightarrow M_{e}\in \mathcal{M}_{E_{e}}$ for an $E_{e}\in 
\mathcal{E}$.

$c_{3}:L\longrightarrow \mathcal{E},a\longrightarrow E_{a}$.

Moreover, $\mu (p,e,q)$ has no \ equivalent in our framework: nevertheless,
if $M_{e}$\ is a first kind measurement, then $q$ can be identified with $%
t_{E_{e}}(S)$ and $\mu (p,e,q)$ with $P_{S}(E_{e}).$

Finally, $c_{1}$ and $c_{3}$ imply that $\nu (p,a)$ can be identified with $%
P_{S}(E_{a})$.

By using the correspondences above, the aforementioned (partial )
translation can be obtained, which shows that also in this case there are
strong structural similarities between Aerts' approach and ours.\medskip

\begin{center}
BIBLIOGRAPHY\smallskip
\end{center}

Aerts, D. (1986). A possible explanation for the probabilities of quantum
mechanics. \textit{J. Math. Physics} \textbf{27}, 202-210.

Aerts, D. (1999). Foundations of quantum physics: a general realistic and
operational approach. \textit{Int. J. Theor. Phys.} \textbf{38}, 289-358.

Aerts, D. and Gabora, L. (2005). A state-context-property model of concepts
and their combinations ii: A Hilbert space representation. \textit{Kibernetes%
} \textbf{34}, 176-204.

Aerts, D. and Sassoli de Bianchi, M. (2014). The extended Bloch
representation of quantum mechanics and the hidden-measurement solution of
the measurement problem. \textit{Ann. Phys}. \textbf{351}, 975-1025.

Aerts, D. and Sassoli de Bianchi, M. (2017). \textit{Universal Measurements.
How to Free Three Birds in One Move.} World Scientific, Singapore.

Aerts, D., Sassoli de Bianchi, M. and Sozzo, S. (2016). On the Foundations
of the Brussels Operational-Realistic Approach to Cognition. \textit{%
Frontiers in Physics}, doi:10.3389/fphy.2016.00017.

Aerts, D., Sozzo, S. and Veloz, T. (2015). Quantum Structures in Cognition
and the Foundations of Human Reasoning. \textit{Int. J. Theor. Phys.} 
\textbf{54}, 4557-4569.

Ballentine, L.E.(1970). The statistical interpretation of quantum mechanics. 
\textit{Rev. Mod. Phys.} \textbf{42}, 368-381.,

Bell, J.S. (1964). On the Einstein-Podolski-Rosen Paradox. \textit{Physics} 
\textbf{1}, 195--200.

Bell, J.S. (1966). On the Problem of Hidden Variables in Quantum Mechanics. 
\textit{Rev. Mod. Phys.} \textbf{38}, 447--452.

Beltrametti, E. and Cassinelli, G. (1981). \textit{The Logic of Quantum
Mechanics.} Reading (MA), Addison-Wesley.

Birkhoff, G. and von Neumann, J. (1936). The Logic of Quantum Mechanics. 
\textit{Ann. Math.} \textbf{37}, 823--843.

Bohr, N. (1958). \textit{Atomic Physics and Human Knowledge}. John Wiley and
Sons, London.

Braithwaite, R.B. (1953). \textit{Scientific Explanation}. Cambridge
University Press, Cambridge.

Busch, P., Lahti, P.J. and Mittelstaedt, P. (1996). \textit{The Quantum
Theory of Measurement}. Springer, Berlin.

Carnap, R. (1966). \textit{Philosophical Foundations of Physics}. Basic
Books Inc., New York.

Dalla Chiara, M. L., Giuntini, R. and Greechie, R. (2004). \textit{Reasoning
in Quantum Theory}. Kluwer, Dordrecht.

Einstein, A., Podolski, B. and Rosen, N. (1935). Can quantum mechanical
description of physical reality be considered complete? \textit{Phys. Rev.} 
\textbf{47}, 777-780.

Feyerabend, F. (1975). \textit{Against Method: Outline of an Anarchist
Theory of Knowledge}. New Left Books, London.

Garola, C. (1999). Semantic realism: a new philosophy for quantum physics. 
\textit{Int. J. Theor. Phys.} \textbf{38}, 3241-3252.

Garola, C. (2008). Physical propositions and quantum languages. \textit{Int.
J. Theor. Phys.} \textbf{47}, 90-103.

Garola, C. (2015). A survey of the ESR model for an objective interpretation
of quantum mechanics. \textit{Int. J. Theor. Phys.} \textbf{54}, 4410-4422.

Garola, C. (2017). Interpreting quantum logic as a pragmatic structure, 
\textit{Int. J. Theor. Phys.} \textbf{56}, 3770-3782.

Garola, C. (2018). An epistemic interpretation of quantum probability via
contextuality. \textit{Found. Sci.,} DOI: 10.1007/s10699-018-9560-4.

Garola, C. and Pykacz, J. (2004). Locality and measurement within the SR
model for an objective interpretation of quantum mechanics. \textit{Found.
Phys.} \textbf{34}, 449-475.

Garola, C. and Persano, M. (2014). Embedding quantum mechanics into a
broader noncontextual theory. \textit{Found. Sci.} \textbf{19}, 217-239.

Garola, C. and Sozzo, S. (2010). Realistic aspects in the standard
interpretation of quantum mechanics. \textit{Humana.ment. J. Phil. Stud.} 
\textbf{13}, 81-101.

Garola, C. and Sozzo, S. (2013). Recovering quantum logic within an extended
classical framework. \textit{Erkenntnis} \textbf{78}, 399-314.

Garola, C., Sozzo, S. and Wu, J. (2016). Outline of a generalization and a
reinterpretation of quantum mechanics recovering objectivity. \textit{Int.
J. Theor. Phys.} \textbf{55}, 2500-2528.

Greenberger, D.M., Horne, M.A., Shimony, A. and Zeilinger, A. (1990). Bell's
theorem without inequalities. \textit{Am. J. Phys.} \textbf{58}, 1131-1143.

Heisenberg, W. (1958). Physics and Philosophy: the Revolution of Modern
Science. Harper, New York.

Hempel, C.C. (1965). \textit{Aspects of Scientific Explanation}. Free Press,
New York.

Kochen, S. and Specker, E. P. (1967). The Problem of Hidden Variables in
Quantum Mechanics. \textit{J. Math. Mech.} \textbf{17}, 59--87.

Khrennikov, A. (2009a). \textit{Contextual Approach to Quantum Formalism}.
Sprin- ger, New York.

Khrennikov, A. (2009b). \textit{Interpretations of Probability}. Walter de
Gruyter, Berlin.

Khrennikov, A. (2015). CHSH inequality: quantum probabilities as classical
conditional probabilities. \textit{Found. Phys. }\textbf{45}, 711-725.

Kuhn, T.S. (1962). \textit{The Structure of Scientific Revolution}. Chicago
University Press, Chicago.

Ludwig, G. (1983). \textit{Foundations of Quantum Mechanics I}. Springer,
New York.

Mermin, N.D. (1993). Hidden variables and the two theorems of John Bell. 
\textit{Rev. Mod. Phys.} \textbf{65}, 803-815.

Nagel, E. (1961). \textit{The Structure of Science}. Harcourt, Brace \&
World, New York.

Piron, C. (1976). \textit{Foundations of Quantum Physics}. Benjamin, Reading
(MA).

R\'{e}dei, N. (1998). \textit{Quantum Logic in Algebraic Approach.} Kluwer,
Dordrecht.

Williamson, J. (2002). Probability logic. In D.B. Gabbay, R.H. Johnson, H.J.
Ohlbach, J. Woods (eds.). \textit{Handbook of the Logic of Argument and
Inference}, I. North-Holland, Amsterdam.

\end{document}